\PassOptionsToPackage{svgnames}{xcolor}
\PassOptionsToPackage{numbers,sort&compress}{natbib}
\documentclass[conference]{IEEEtran}
\IEEEoverridecommandlockouts

\usepackage{tikz}
\usepackage{amsmath}
\usepackage{enumitem}

\usepackage{booktabs}
\usepackage{adjustbox}
\usepackage{tabularx}
\usepackage{url}

\usepackage{wrapfig}
\usepackage{hyperref}

\usepackage[switch]{lineno}

\usepackage{import}
\ifx\homepath\overleafhome
  
\usepackage{etoolbox}

\newtoggle{AlgsNoEnd}

\newif\iftr     % Full arXiv technical report
\newif\ifconf   % Size-constrainted Submission to a conf or journal
\newif\ifnonb   % Non blind submission

\usepackage[utf8]{inputenc}
\newcommand{\code}[1]{\texttt{#1}}
\usepackage[binary-units=true]{siunitx}
\usepackage{xspace}
\usepackage[svgnames]{xcolor}

\usepackage{collab}
\usepackage[numbers,sort&compress]{natbib}

\usepackage{soul}
\soulregister\cite7
\soulregister\autoref7
\soulregister\code7
\soulregister\toolname7
\soulregister\ref7
\soulregister\pageref7
\usepackage{hhline}
\definecolor{lightyellow}{RGB}{250, 250, 180}
\definecolor{HLYELLOW}{RGB}{240, 127, 0}
\definecolor{hlyellow}{RGB}{240, 127, 0}
\sethlcolor{lightyellow}

\usepackage{algorithm}
\usepackage{algpseudocode}
\usepackage{pifont}
\usepackage{amsmath}
\algdef{SE}[FOREACHIN]{ForEachIn}{EndForEachIn}[2]{\algorithmicfor\ \textbf{each}\ #1\ \textbf{in}\ #2\ \textbf{do}}{\algorithmicend\ \algorithmicfor}%
\algdef{SE}[FOREACHP]{ForEachP}{EndForEachP}[1]{\algorithmicfor\ \textbf{each}\ #1\ \textbf{in parallel do}}{\algorithmicend\ \algorithmicfor}%
\algdef{SE}[FOREACHINP]{ForEachInP}{EndForEachInP}[2]{\algorithmicfor\ \textbf{each}\ #1\ \textbf{in}\ #2\ \textbf{in parallel do}}{\algorithmicend\ \algorithmicfor}%
\algdef{SE}[FORMOD]{ForMod}{EndForMod}[3]{\algorithmicfor\ #1\ \textbf{from}\ #2\ \textbf{to}\ #3\ \textbf{do}}{\algorithmicend\ \algorithmicfor}%
\algdef{SE}[FORMOD]{ForModS}{EndForModS}[4]{\algorithmicfor\ #1\ \textbf{from}\ #2\ \textbf{to}\ #3\ \textbf{step}\ #4\ \textbf{do}}{\algorithmicend\ \algorithmicfor}%
\algdef{SE}[FORMODP]{ForModPS}{EndForModPS}[3]{\algorithmicfor\ #1\ \textbf{from}\ #2\ \textbf{to}\ #3\ \textbf{in parallel do}}{\algorithmicend\ \algorithmicfor}%
\algdef{SE}[FORMODP]{ForModP}{EndForModP}[4]{\algorithmicfor\ #1\ \textbf{from}\ #2\ \textbf{to}\ #3\ \textbf{step} #4\ \textbf{in parallel do}}{\algorithmicend\ \algorithmicfor}%

\iftoggle{AlgsNoEnd} {
  \algtext*{EndFunction}% Remove "end while" text
  \algtext*{EndIf}% Remove "end if" text
  \algtext*{EndFor}% Remove "end if" text
  \algtext*{EndForEachIn}% Remove "end if" text
  \algtext*{EndForEachP}% Remove "end if" text
  \algtext*{EndForEachInP}% Remove "end if" text
  \algtext*{EndForModP}% Remove "end if" text
  \algtext*{EndForModPS}% Remove "end if" text
  \algtext*{EndForModS}% Remove "end if" text
}{}

\algdef{SE}[VARIABLES]{Variables}{EndVariables}{\algorithmicvariables}
  {\algorithmicend\ \algorithmicvariables}
  \algnewcommand{\algorithmicvariables}{\textbf{global}}
\algnewcommand{\LineComment}[1]{\State \(\triangleright\) #1}
\algnewcommand{\And}{\textbf{and}\xspace}

\usepackage{tikz}
\usepackage{pifont}
\usetikzlibrary{shapes}

\usepackage{listings}
\newfloat{codeblock}{H}{myc}
\definecolor{darkblue}{rgb}{0,0,.6}
\definecolor{darkred}{rgb}{.6,0,0}
\definecolor{darkgreen}{rgb}{0,.5,0}
\definecolor{red}{rgb}{.98,0,0}
\definecolor{gray}{rgb}{.6,.6,.6}
\definecolor{newgreen}{RGB}{169,209,142}
\definecolor{newpurple}{RGB}{237,134,254}
\definecolor{neworange}{RGB}{244,177,131}
\definecolor{newyellow}{RGB}{255,217,102}
\else
  
\usepackage{etoolbox}

\newtoggle{AlgsNoEnd}

\newif\iftr     % Full arXiv technical report
\newif\ifconf   % Size-constrainted Submission to a conf or journal
\newif\ifnonb   % Non blind submission

\usepackage[utf8]{inputenc}
\newcommand{\code}[1]{\texttt{#1}}
\usepackage[binary-units=true]{siunitx}
\usepackage{xspace}
\usepackage[svgnames]{xcolor}

\usepackage{collab}
\usepackage[numbers,sort&compress]{natbib}

\usepackage{soul}
\soulregister\cite7
\soulregister\autoref7
\soulregister\code7
\soulregister\toolname7
\soulregister\ref7
\soulregister\pageref7
\usepackage{hhline}
\definecolor{lightyellow}{RGB}{250, 250, 180}
\definecolor{HLYELLOW}{RGB}{240, 127, 0}
\definecolor{hlyellow}{RGB}{240, 127, 0}
\sethlcolor{lightyellow}

\usepackage{algorithm}
\usepackage{algpseudocode}
\usepackage{pifont}
\usepackage{amsmath}
\algdef{SE}[FOREACHIN]{ForEachIn}{EndForEachIn}[2]{\algorithmicfor\ \textbf{each}\ #1\ \textbf{in}\ #2\ \textbf{do}}{\algorithmicend\ \algorithmicfor}%
\algdef{SE}[FOREACHP]{ForEachP}{EndForEachP}[1]{\algorithmicfor\ \textbf{each}\ #1\ \textbf{in parallel do}}{\algorithmicend\ \algorithmicfor}%
\algdef{SE}[FOREACHINP]{ForEachInP}{EndForEachInP}[2]{\algorithmicfor\ \textbf{each}\ #1\ \textbf{in}\ #2\ \textbf{in parallel do}}{\algorithmicend\ \algorithmicfor}%
\algdef{SE}[FORMOD]{ForMod}{EndForMod}[3]{\algorithmicfor\ #1\ \textbf{from}\ #2\ \textbf{to}\ #3\ \textbf{do}}{\algorithmicend\ \algorithmicfor}%
\algdef{SE}[FORMOD]{ForModS}{EndForModS}[4]{\algorithmicfor\ #1\ \textbf{from}\ #2\ \textbf{to}\ #3\ \textbf{step}\ #4\ \textbf{do}}{\algorithmicend\ \algorithmicfor}%
\algdef{SE}[FORMODP]{ForModPS}{EndForModPS}[3]{\algorithmicfor\ #1\ \textbf{from}\ #2\ \textbf{to}\ #3\ \textbf{in parallel do}}{\algorithmicend\ \algorithmicfor}%
\algdef{SE}[FORMODP]{ForModP}{EndForModP}[4]{\algorithmicfor\ #1\ \textbf{from}\ #2\ \textbf{to}\ #3\ \textbf{step} #4\ \textbf{in parallel do}}{\algorithmicend\ \algorithmicfor}%

\iftoggle{AlgsNoEnd} {
  \algtext*{EndFunction}% Remove "end while" text
  \algtext*{EndIf}% Remove "end if" text
  \algtext*{EndFor}% Remove "end if" text
  \algtext*{EndForEachIn}% Remove "end if" text
  \algtext*{EndForEachP}% Remove "end if" text
  \algtext*{EndForEachInP}% Remove "end if" text
  \algtext*{EndForModP}% Remove "end if" text
  \algtext*{EndForModPS}% Remove "end if" text
  \algtext*{EndForModS}% Remove "end if" text
}{}

\algdef{SE}[VARIABLES]{Variables}{EndVariables}{\algorithmicvariables}
  {\algorithmicend\ \algorithmicvariables}
  \algnewcommand{\algorithmicvariables}{\textbf{global}}
\algnewcommand{\LineComment}[1]{\State \(\triangleright\) #1}
\algnewcommand{\And}{\textbf{and}\xspace}

\usepackage{tikz}
\usepackage{pifont}
\usetikzlibrary{shapes}

\usepackage{listings}
\newfloat{codeblock}{H}{myc}
\definecolor{darkblue}{rgb}{0,0,.6}
\definecolor{darkred}{rgb}{.6,0,0}
\definecolor{darkgreen}{rgb}{0,.5,0}
\definecolor{red}{rgb}{.98,0,0}
\definecolor{gray}{rgb}{.6,.6,.6}
\definecolor{newgreen}{RGB}{169,209,142}
\definecolor{newpurple}{RGB}{237,134,254}
\definecolor{neworange}{RGB}{244,177,131}
\definecolor{newyellow}{RGB}{255,217,102}
\fi

\newcommand{\toolname}{\emph{UPM}\xspace}
\collabAuthor{copik}{blue}{Marcin}
\collabAuthor{lex}{purple}{Alex}
\collabAuthor{todos}{red}{TODO}
\collabAuthor{wei}{darkgreen}{Wei}
\collabAuthor{htor}{magenta}{Torsten}

\usepackage{caption}
\usepackage{subcaption}

\usepackage{moresize}

\usepackage[many]{tcolorbox}
\tcbuselibrary{skins}
\usetikzlibrary{calc}

\definecolor{myblue}{RGB}{0,163,243}
\definecolor{textnewGreen}{HTML}{66AE3E}
\definecolor{bgnewGreen}{HTML}{EBF4DE}
\newtcbox{\titleboxgreen}{
  enhanced,
  colupper=white,
 colback=textnewGreen,
  fontupper=\bfseries\sffamily\large,
  size=small,
  baseline=4pt,
  nobeforeafter,
	left=0pt,
	right=0pt,
	top=0pt,
	bottom=-1pt,
  frame code={
    \path[fill=textnewGreen] (frame.north west)
    -- ([xshift=2mm]frame.north east)
    -- (frame.south east)
    -- (frame.south west)
    -- (frame.north west)
      [sharp corners]-- cycle;
  }
}

\makeatletter
\newtcolorbox{summarygreen}[1]{
  enhanced,
  skin=bicolor,
  arc=0pt,
	left=0pt,
	right=0pt,
	top=0pt,
	bottom=0pt,
  coltitle=white,
  colframe=textnewGreen,
  colback=bgnewGreen,
  colbacklower=white,
  detach title,
  title={#1}
}
\makeatother

\begin{document}
%-------------------------------------------------------------------------------

%don't want date printed
%\date{}

\title{User-guided Page Merging for Memory Deduplication in Serverless Systems}

\IEEEoverridecommandlockouts

\author{
    \IEEEauthorblockN{
        Wei Qiu\textsuperscript{1}\IEEEauthorrefmark{1},
        Marcin Copik\IEEEauthorrefmark{2},
        Yun Wang\textsuperscript{2}\IEEEauthorrefmark{3},
        Alexandru Calotoiu\IEEEauthorrefmark{2},
        Torsten Hoefler\IEEEauthorrefmark{2}
    }
    \IEEEauthorblockA{\IEEEauthorrefmark{1}Huawei Technologies, China}
    \IEEEauthorblockA{\IEEEauthorrefmark{2}Department of Computer Science,
      ETH Z{\"u}rich, Z{\"u}rich, Switzerland}
    \IEEEauthorblockA{\IEEEauthorrefmark{3}Shanghai Jiao Tong University, China}
    %\\Email: \{marcin.copik, alexandru.calotoiu, htor\}@inf.ethz.ch}
    %\IEEEauthorblockA{\IEEEauthorrefmark{2}Microsoft
    %\\Email: kotaranov@microsoft.com}
    %\IEEEauthorblockA{\IEEEauthorrefmark{2}Tencent Technology}
    \IEEEauthorrefmark{2}firstname.lastname@inf.ethz.ch
    %\\Email: \{marcin.copik, alexandru.calotoiu, htor\}@inf.ethz.ch}
    %\\Email: kotaranov@microsoft.com}
}

%Work on this paper was done while at Tencent Technology.
%\IEEEpubid{\makebox[\columnwidth]{979-8-3503-2445-7/23/\$31.00 ©2023\hfill} \hspace{\columnsep}\makebox[\columnwidth]{ }}

\maketitle
\setcounter{footnote}{1}
\footnotetext{Work on this paper was done while at ETH Z{\"u}rich.}
\setcounter{footnote}{2}
\footnotetext{Work on this paper was done while at Tencent Technology.}

\IEEEpubidadjcol

%-------------------------------------------------------------------------------
\begin{abstract}
%-------------------------------------------------------------------------------

Serverless computing is an emerging cloud paradigm that offers an elastic and scalable allocation of computing resources with pay-as-you-go billing.
In the Function-as-a-Service (FaaS) programming model, applications comprise short-lived and stateless serverless functions executed in isolated containers or microVMs, which can quickly scale to thousands of instances and process terabytes of data.
This flexibility comes at the cost of duplicated runtimes, libraries, and user data spread across many function instances, and cloud providers do not utilize this redundancy.
The memory footprint of serverless forces removing idle containers to make space for new ones, which decreases performance through more cold starts and fewer data caching opportunities.

We address this issue by proposing deduplicating memory pages of serverless workers with identical content, based on the content-based page-sharing concept of Linux Kernel Same-page Merging (KSM).
We replace the background memory scanning process of KSM, as it is too slow to locate sharing candidates in short-lived functions.
Instead, we design User-Guided Page Merging (UPM), a built-in Linux kernel module that leverages the \texttt{madvise} system call: we enable users to advise the kernel of memory areas that can be shared with others.
We show that UPM reduces memory consumption by up to 55\% on 16 concurrent containers executing a typical image recognition function, more than doubling the density for containers of the same function that can run on a system.
\end{abstract}

\noindent \textbf{Implementation}: \url{https://github.com/spcl/UPM}

%-------------------------------------------------------------------------------
\section{Introduction}

Serverless computing allows users to deploy elastic and scalable applications decomposed into fine-grained and stateless functions, at the cost of increased overheads and duplicated resources.
\iftr
Deploying to FaaS allows developers to delegate operational concerns such as resource provisioning, maintenance, and fault tolerance to the cloud provider~\cite{currenttrend},
\else
Deploying to FaaS allows developers to delegate operational concerns such as resource provisioning, maintenance, and fault tolerance to the cloud provider,
\fi
and it has been used in parallel and compute-intensive tasks such as big data analytics, machine learning, and high-performance computing~\cite{227653,10.1145/3318464.3389758,9671899,2021rfaas,copik2022fmi}.
In FaaS, functions are deployed in isolated sandboxes for security reasons on dynamically allocated resources, 
and they can be scaled down to zero when traffic in the system decreases.
While this operational model provides cloud operators with flexible and efficient allocation and scheduling,
it leads to resource waste caused by duplicated function environments for each function instance~\cite{photons,10.1145/3492321.3524272,10.1145/3445814.3446714}.
%
%In particular, each sandbox duplicates language runtime, libraries, and function data.
%
%functions runtimes need to be changed to share [photon, medes, faaasm]

%Memory consumption is a major constraint that prevents allocating m.
Reducing memory pressure has many benefits for both the cloud operator and users.
\iftr
When function sandboxes can be co-located on the same host machine, resource utilization and data sharing through caches are increased~\cite{10.5555/3277355.3277369,10.1145/3445814.3446757,romero2021faat,10.1145/3447786.3456239},
and performance of data analytics is increased~\cite{servermix}.
Exhausting available memory forces operators to remove idle and warm containers, increasing the rate of cold startups,
a major issue in serverless computing~\cite{sebs,Manner2018ColdSI}.
\else
When function sandboxes can be co-located on the same host machine, resource utilization and data sharing through caches are increased~\cite{10.5555/3277355.3277369,10.1145/3447786.3456239},
and performance of data analytics is increased~\cite{servermix}.
%In addition, the co-location creates an opportunity to reduce the memory footprint of
%functions storing in the memory contents of the same libraries and data objects.
%
Exhausting available memory forces operators to remove idle and warm containers, increasing the rate of cold startups,
a major issue in serverless computing~\cite{sebs}.
\fi
%Cloud providers would benefit from decreasing memory pressure, as it would allow them
%to increase container retainment time and reduce the rate of cold startups, a major issue
%in serverless computing~\cite{sebs,Manner2018ColdSI}.
%
\iftr
Furthermore, reducing bloat increases the memory pool for low-latency storage needed in stateful serverless~\cite{faaslets,10.14778/3407790.3407836,praas}.
\else
Furthermore, reducing bloat increases the memory pool for low-latency storage needed in stateful serverless~\cite{faaslets,10.14778/3407790.3407836}.
\fi
Leaner platforms can host more functions with larger memory allocations, which
allows serverless to process big data workloads faster and cheaper~\cite{227653}.

%While cloud operators oversubscribe resources and host many invocations on the same machine, the bloated function memory footprint limits server sharing and decreases system efficiency.

%Memory consumption is a major constraint that prevents allocating many function sandboxes on the same machine.
%
The landscape of memory consumption in serverless varies significantly, with 50\% and 90\% of different functions consuming at most 170 MB and 400 MB of memory, respectively~\cite{254430}.
A prevalent use case is machine learning inference~\cite{ali2020batch,10.1145/3472456.3472501,234998},
characterized by substantial computational requirements and large memory allocations that can require upwards of 500 MB of memory per instance~\cite{sebs,9219119}.
The trained model significantly contributes to this memory consumption, accounting for a major portion of process memory identical across functions (Sec.~\ref{chapter:profiling}).
\iftr
While deep learning inference has been moving to the edge~\cite{8675201,8876870,Xu2018}, not all mobile and edge devices can support computationally intensive inference on large models~\cite{8675201,8876870,Xu2018,8736011}.
Serverless functions could be a perfect platform for implementing
the required edge-cloud cooperation~\cite{9139674,10.1145/3431379.3460646,8567674,9474932}.
\else
While deep learning inference has been moving to the edge, not all mobile and edge devices can support computationally intensive inference on large models~\cite{8675201,8876870}.
Serverless functions could be a perfect platform for implementing
the required edge-cloud cooperation~\cite{9139674,10.1145/3431379.3460646,9474932}.
\fi
The growing interest in deploying machine learning in cloud functions, together with the increasing model size, requires
new solutions to reduce memory redundancy and prevent resource exhaustion when scaling isolated and containerized function deployments.
%
%Since concurrent invocations of the same function are common~\cite{photons}, 

%Serverless 
%~\cite{photons}

%Just like functions reclaim idle resources from virtual machines, deduplication reclaims idle memory from workers.

Unfortunately, current memory sharing functionalities are insufficient for deduplication in serverless functions.
\iftr
Techniques such as Kernel Same-page Merging (KSM) are not designed to handle volatile and short-term workloads as they can require over half an hour to locate deduplication opportunities~\cite{xen,202637} (Sec.~\ref{chapter:related}).
\else
Techniques such as Kernel Same-page Merging (KSM) are not designed to handle volatile and short-term workloads as they can require over half an hour to locate deduplication opportunities~\cite{202637} (Sec.~\ref{chapter:related}).
\fi
On the serverless side, systems adopt the ideas of caching and checkpointing function sandboxes~\cite{SAND,SOCK,replayable}, co-locating instances of the same function within a single process~\cite{photons,faaslets}, and decomposing containers into deduplicated state~\cite{10.1145/3492321.3524272}.
Such approaches are limited by sharing memory only between instances of the same function,
explicit programming of shared memory objects, and deep changes to the entire serverless software stack.
%and requiring changes to function code or 
%or requiring users to access shared data explicitly, or requchanging the entire sandbox model.
%
%A serverless memory deduplication technique should be agnostic of a function's language and cloud runtime,
%and it should not require deep modifications of the entire software stack to enable memory sharing.

To address this issue, we propose \textbf{User-guided Page Merging (UPM)},
a built-in Linux kernel module that enables page sharing across
different functions with an affordable time overhead.
We extend the semantics of Kernel Same-page Merging by replacing the background memory scanning process with an
on-demand, user-guided deduplication that better suits serverless workloads.
\iftr
We use the existing \texttt{madvise}~\cite{madvise} Linux system call to implement the interaction between the user function and the deduplication running in kernel space.
\else
We use the existing \texttt{madvise} Linux system call to implement the interaction between the user function and the deduplication running in kernel space.
\fi
The \emph{madvised} memory pages are treated as potentially mergeable regions.
They are shared in a copy-on-write fashion if any page with identical content has been \emph{madvised} before.

\toolname{} is designed to be simple, flexible, low overhead, and easy to use.
\toolname{} is agnostic of language, runtime, and containers the FaaS system uses.
In \toolname{}, deduplication is applied only once on the cold container, and consecutive warm invocations benefit from page sharing without CPU overheads.
Furthermore, \toolname{} extends the security of memory sharing through KSM and page caches by introducing controlled, \emph{opt-in} sharing.
Using \toolname{} is very simple and seamless, as users only have to annotate shareable memory regions,
%and manual data management with shared memory~\cite{faaslets,praas} is unnecessary.
and manual data management with shared memory~\cite{faaslets} is unnecessary.
%because the operating system seamlessly shares the memory pages.
%
Our technique works for both file-backed and anonymous memory and enables sharing between different functions.
%as long as they have memory pages with the same content.
%
We show experimentally that \toolname{} successfully reduces the system's memory utilization by up to 55\% when 16 concurrent containers are deployed.

We make the following contributions:
\begin{itemize}
  \item We provide a thorough and detailed study on the memory sharing potential of serverless workloads (Sec.~\ref{chapter:profiling}).
  %\item We propose \toolname{}, a novel methodology for memory deduplication across serverless functions, and we implement \toolname{} as a built-in module of the Linux kernel (Sec.~\ref{chapter:implementation}).
  \item We propose \toolname{}, a novel methodology for memory deduplication across serverless functions, and we implement \toolname{} as a built-in module of Linux kernel (Sec.~\ref{chapter:implementation}).
  %\item We demonstrate empirically that \toolname{} reduces up to 55\% of memory consumption in a machine learning case study (Sec.~\ref{chapter:evaluation}).
  \item We demonstrate that \toolname{} reduces up to 55\% of memory consumption in a machine learning case study (Sec.~\ref{chapter:evaluation}).
\end{itemize}

%-------------------------------------------------------------------------------

%-------------------------------------------------------------------------------
\section{Background and Related Work}

\label{chapter:related}

Memory deduplication has been employed in datacenters for decades.
However, such systems need several minutes to discover deduplication opportunities and are thus limited to long-running and stationary workloads.
%
%Serverless workloads are characterized by short-running invocations and thus require a different approach
%that can overcome this limitation.
Serverless workloads require a different approach to support short-running invocations.
While optimized serverless runtimes can provide memory sharing across invocations, they do not have the same level of generality offered by deduplication techniques.

\subsection{Memory Deduplication}
Page sharing has been proposed %in the Disco system
to share memory pages of files accessed by different virtual machines~\cite{10.1145/265924.265930}.
\iftr
Content-based page sharing has been first implemented in VMware ESX~\cite{vmware}, and later 
in the Linux kernel as the Kernel Same-page Merging (KSM)~\cite{ksm_paper, 6799096}.
\else
Content-based page sharing has been implemented in the Linux kernel as the Kernel Same-page Merging (KSM)~\cite{ksm_paper}.
\fi
There, memory pages are scanned and hashed to determine pages in guest virtual machines with the same content.
When the hash value is already present in the system, a byte-by-byte comparison is conducted to guarantee
that pages are truly identical and their contents have not changed.
Then, pages are merged into one physical page in a copy-on-write manner, regardless of origin.
%
%An alternative approach requires modifying paravirtualized I/O to scan data read from disk to detect duplicated content
%in file-backed data and page caches~\cite{202637}.
Other approaches modify paravirtualized I/O to detect duplicated content in file-backed data and page caches~\cite{202637}.
%An alternative approach requires modifying paravirtualized I/O to scan data read from disk to detect duplicated content.
%
%This approach is motivated by most memory savings coming from file-backed data and page caches~\cite{202637}.
%
The deduplication is most efficient on mixed CPU and I/O workloads,
reducing memory footprint by up to 60\%~\cite{5951913,vmware,10.1145/1618525.1618529}.

%Page sharing has been proposed in the Disco system to share memory pages of files accessed by different virtual machines~\cite{10.1145/265924.265930}.
%
%Content-based page sharing has been first implemented in VMware ESX~\cite{vmware}.
%
%In this technique, memory pages are scanned and hashed to determine pages in guest virtual machines with the same content.
%%
%When the hash value has already been found in the system, a byte-by-byte comparison is conducted to guarantee
%that pages are truly identical and their contents have not changed since hash computation.
%
%Then, pages are merged into one physical page in a copy-on-write manner, regardless of their origin.
%
%An alternative approach requires modifying paravirtualized I/O to scan data read from disk to detect duplicated content.
%
%This approach is motivated by most memory savings coming from file-backed data and page caches~\cite{202637}.
%
%The deduplication is most efficient on mixed CPU and I/O workloads,
%and it can reduce memory footprint by up to 60\%~\cite{5951913,vmware,10.1145/1618525.1618529}.
%
%Page sharing has been implemented in the Xen~\cite{xen} hypervisor and the Linux kernel as the Kernel Same-page Merging (KSM)~\cite{ksm_paper, ksm_other}.

\paragraph{Limitations}
Memory pages are scanned periodically to keep the CPU overhead below 10-20\%~\cite{Sharma2012},
and increasing deduplication efficiency requires tolerating higher CPU overheads~\cite{6799096}, up to 68\% when
scanning every 20 milliseconds~\cite{179045}.
\iftr
On the other hand, keeping the compute overhead moderate means that the discovery of duplicated contents can take as long as 40 minutes and pages alive for less than 5 minutes cannot be deduplicated~\cite{202637}, as scanning just 50 megabytes requires over two minutes in the KSM~\cite{amt-ksm}.
\else
On the other hand, keeping the compute overhead moderate means that discovery of duplicated contents can take as long as 40 minutes~\cite{202637,xen} and pages alive for less than 5 minutes cannot be deduplicated~\cite{202637}, as scanning just 50 megabytes requires over two minutes in the KSM~\cite{amt-ksm}.
\fi
In the serverless world, where the median and 90th percentile of function duration are about 3 and 60 seconds~\cite{254430}, respectively,
functions will complete execution before the system finds sharing candidates of the function's memory.
On AWS Lambda, function containers can be scaled down after only 6.5 minutes of no activity~\cite{sebs}, which
is not enough to discover deduplication opportunities.
Furthermore, even if a function container is used frequently and retained long enough to benefit from page sharing,
it would still handle many function invocations with bloated memory.

\paragraph{Optimized Sharing}
Deduplication can be optimized with I/O-based hints~\cite{179045}
and page classification~\cite{10.1145/2674025.2576204}.
Still, the time needed for deduplication is in the order of minutes.
Difference Engine~\cite{difference_engine} enables sub-page sharing, i.e., deduplicating pages that are not completely identical, by constructing patches containing the difference relative to a reference page.
Such systems can outperform other deduplication techniques by locating pages with just a few bytes of difference
that would otherwise never be shared.
Similarly, virtual machines can be classified according to the similarity of memory contents or their software stack~\cite{10.1145/1989493.1989554,10.1145/1618525.1618529}, 
motivating co-location of machines recognized as similar.
%
%The address space layout randomization (ASLR)~\cite{pax_pre, pax} improves systems security, but it hurts the
%effectiveness of memory deduplication~\cite{aslr_effect}.

\paragraph{ASLR}
The address space layout randomization (ASLR) improves systems security, but it hurts the
effectiveness of memory deduplication~\cite{aslr_effect}. 
%
%A study shows that ASLR increases physical memory consumption by more than 18\% on just four virtual machines with KSM enabled.
%
Sharing efficiency can be decreased by ASLR by 10-13 percentage points~\cite{180961},
and by 5\% on FaaS workloads~\cite{10.1145/3492321.3524272}.
While these negative effects can be mitigated~\cite{VANOGARCIA202077}, the effects of ASLR on memory duplication in serverless are not widely studied.

%Difference Engine~\cite{difference_engine} enables sub-page sharing, i.e., deduplication of pages that are not completely identical.
%
%The system achieves that by constructing patches that represent a page as the difference relative to a reference page.
%
%Systems such as Difference Engine can outperform other deduplication techniques by locating pages with just a few bytes of difference that would otherwise never be shared.
%
%Memory Buddies~\cite{10.1145/1618525.1618529} generates fingerprints for virtual machines to predict which machines should have
%a large similarity of memory contents.
%
%Similarly, virtual machines can be classified according to the similarity of memory contents or their software stack~\cite{10.1145/1989493.1989554,10.1145/1618525.1618529}, 
%motivating co-location of machines recognized as similar.
%
%The address space layout randomization (ASLR)~\cite{pax_pre, pax} improves systems security, but it hurts the
%effectiveness of memory deduplication~\cite{aslr_effect}.
%
%A study shows that ASLR increases physical memory consumption by more than 18\% on just four virtual machines with KSM enabled.
%Another study revelead that sharing efficiency can be decreased by ASLR by 10-13 percentage points~\cite{180961}.
%
%While these negative effects can be mitigated~\cite{VANOGARCIA202077}, the effects of ASLR on memory duplication in serverless are unstudied.

\begin{summarygreen}{Summary}
%
%Traditional content-based deduplication techniques are ill-suited for dynamic and short-running functions.
%
The slow convergence of content-based deduplication was appropriate for virtual machines whose lifetime was weeks and months, but it does not fit into serverless containers.
\end{summarygreen}

\subsection{Serverless}
%
%Serverless platforms include servers responsible for allocating and running functions instances.
%
\iftr
Function instances are placed in containers~\cite{234857}, microVMs~\cite{agache2020firecracker}, unikernels~\cite{10.1145/3342195.3392698} and dedicated sandboxes~\cite{catalyzer}.
\else
Function instances are placed in containers, microVMs, unikernels, and dedicated sandboxes.
\fi
The sandbox allows a single server to handle many functions concurrently while preserving isolation guarantees.
Unlike virtual machines requiring deduplication to remove identical files, containers share the page cache.
%
%With the help of virtual file systems such as Docker OverlayFS~\cite{overlayfs}, the same files should have a single copy in memory across many containers.
With the help of virtual file systems such as Docker OverlayFS, the same files should have a single copy in memory across many containers.

%\paragraph{Memory Sharing for Serverless Systems.}
%
To date, several systems have investigated how to reduce the memory consumption of serverless functions.
SAND~\cite{SAND} and SOCK~\cite{SOCK} use caching to reduce the startup time and share the runtime memory.
They use the cache of previous instances for consecutive invocations of a function and then share the memory using copy-on-write, resulting in automatic memory sharing.
Replayable Execution~\cite{replayable} and Catalyzer~\cite{catalyzer} adopt the Checkpoint/Restore (C/R) technique to reuse function memory.
This technique checkpoints the application sandbox state as an image and consecutive invocations are restored from the checkpointed image.
SEUSS~\cite{10.1145/3342195.3392698} provides page sharing when deploying unikernels from a snapshot,
and a similar approach can be used to pre-initialize functions at build time and deploy from an image
heap~\cite{10.1145/3360610}.
Photons~\cite{photons} and Faasm~\cite{faaslets} share the runtime and application state by co-locating multiple
instances of the same function within the same process.
These systems focus on sharing the runtime and the application data across different invocations of the same function. None propose a general solution for memory sharing across \emph{different} functions.
Other techniques target memory sharing for specific applications, such as deep learning inference~\cite{280704}.
%
%Furthermore, none of these techniques are genuinely generic as they depend on specific or dedicated languages, runtimes, and containers, limiting their applicability to the rich world of serverless clouds.

Medes deduplicates memory by introducing a new deduplicated state of serverless containers~\cite{10.1145/3492321.3524272}.
Warm containers are snapshotted, and their memory pages are fingerprinted to find duplicated content.
While this method is independent of the function language, it requires deep changes and major additions in serverless platforms to support this new container state.
Furthermore, it comes with new system components that need to be deployed on each server handling function requests.

\begin{summarygreen}{Summary}
%
%Traditional content-based deduplication techniques are ill-suited for dynamic and short-running functions.
%
Existing serverless runtimes can support memory sharing, but only in the limited case of invocations of the same function.
Other approaches require dedicated runtimes and container systems.
There is a need for an agnostic approach to memory deduplication that supports the rich serverless world of different
languages, runtimes, and sandboxes.
\end{summarygreen}

%-------------------------------------------------------------------------------
\section{Sharing Potential}

\label{chapter:profiling}
%-------------------------------------------------------------------------------

The efficiency of memory deduplication varies depending on the system, employed workloads, and software stack similarity.
For example, self-sharing within a single machine is often the primary source of memory savings and even minor software differences significantly impact inter-machine sharing~\cite{180961}.
On the other hand, deduplication on mobile operating systems is increased by 56\% when considering sub-page sharing on 1kB segments~\cite{lee2015memscope}.
Serverless functions represent a different class of workloads, as their memory footprint is dominated by language runtime and user code, and many operating system components are shared by default.
Therefore, to understand how different deduplication strategies can perform on serverless workloads, we conduct
a profiling study of functions to analyze the sharing potential, shareable memory type, sub-page sharing, and ASLR implications.

\begin{figure}[t!]
    \centering
    \begin{subfigure}{\linewidth}
     %\centering
     \includegraphics[width=\linewidth]{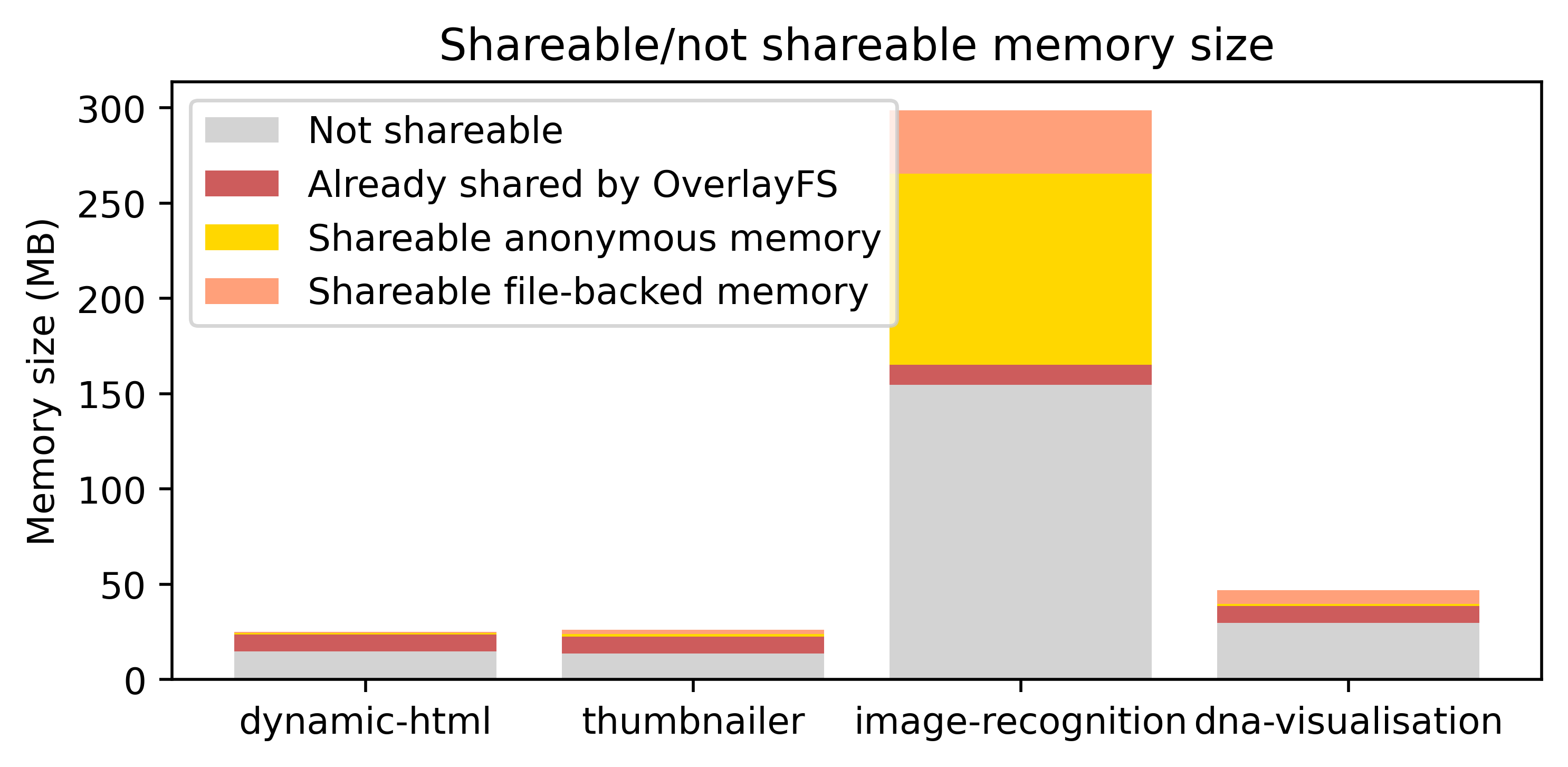}
    %  \caption{Absolute memory size}
     \label{fig:y equals x}
    \end{subfigure}
    \hfill
    \begin{subfigure}{\linewidth}
     %\centering
     \includegraphics[width=\linewidth]{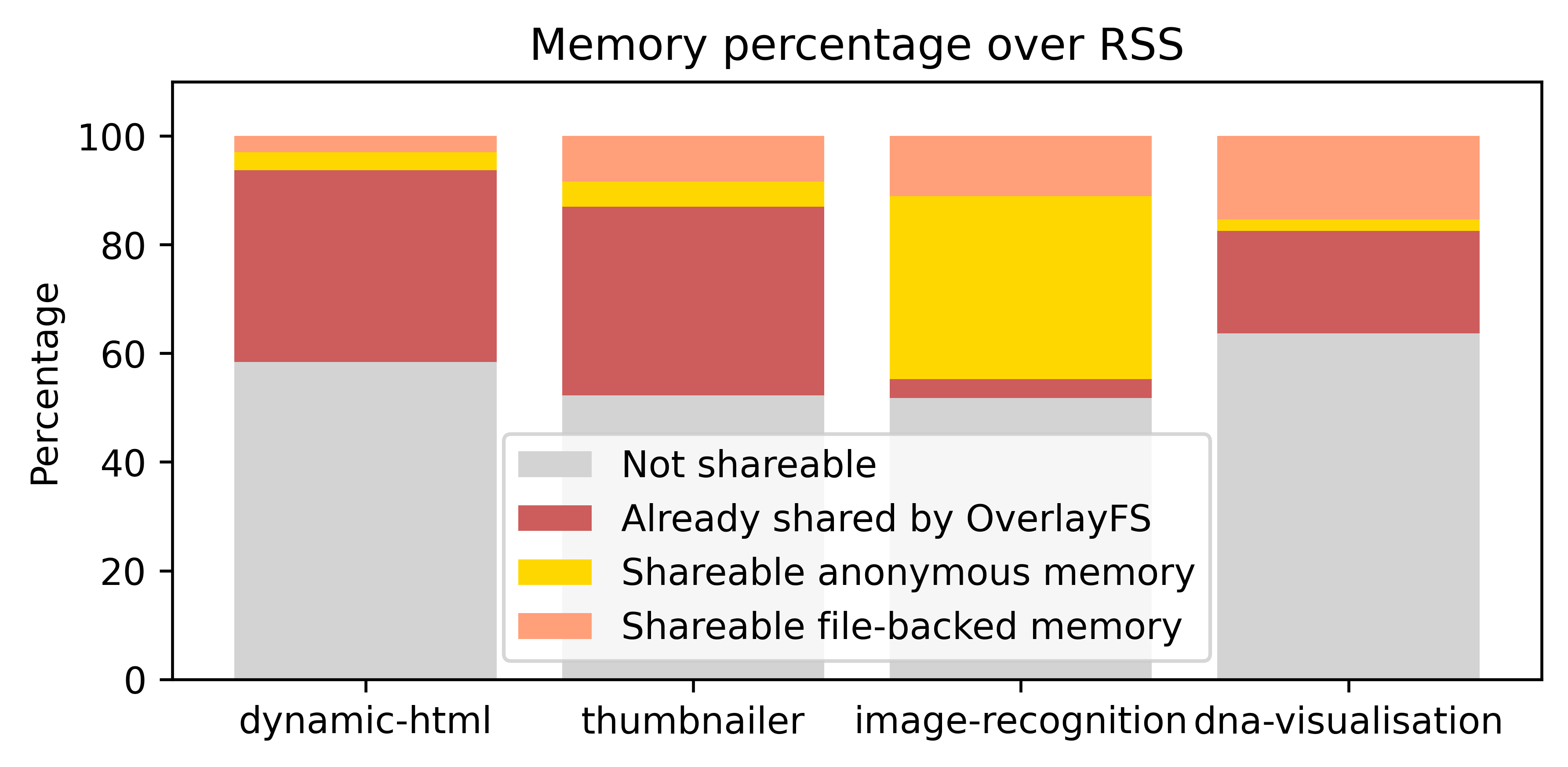}
    %  \caption{Relative memory size over RSS}
     \label{fig:three sin x}
    \end{subfigure}
    \caption{Memory sharing potential in serverless functions.}
    \label{fig:sharing_potential}
\end{figure}

\begin{figure*}[ht!]
     \centering
     \begin{subfigure}{0.39\linewidth}
         \centering
         \includegraphics[width=\linewidth]{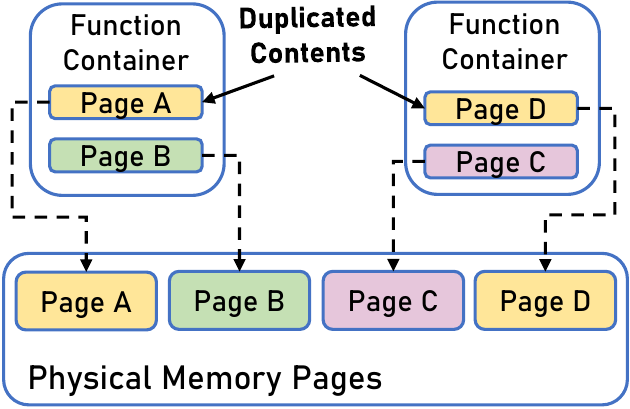}
         \caption{FaaS without \toolname{}.}
         \label{fig:faas without usm}
     \end{subfigure}
     \hfill
     \begin{subfigure}{0.59\textwidth}
         \centering
         \includegraphics[width=\textwidth]{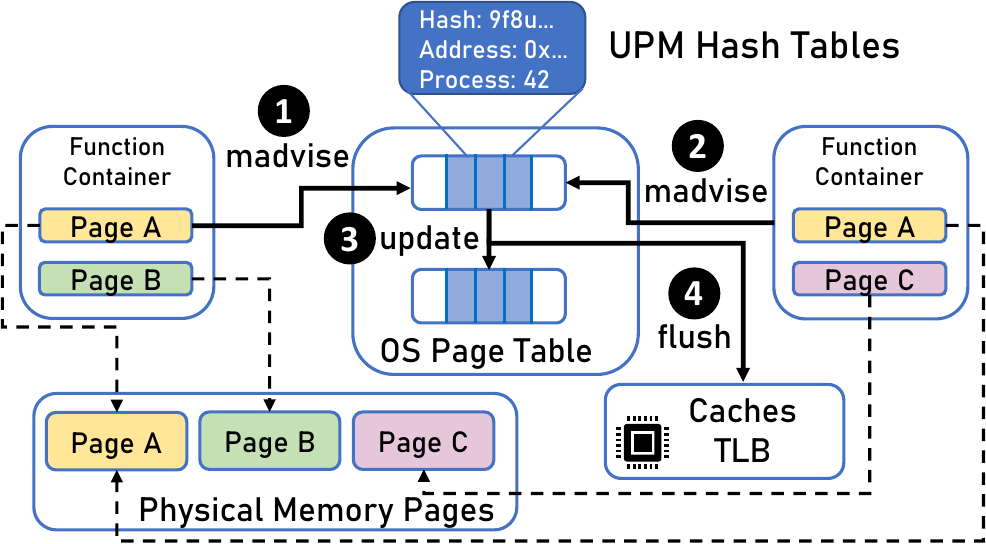}
         \caption{FaaS with \toolname{}.}
         \label{fig:faas with usm}
     \end{subfigure}
    \caption{\textbf{Memory deduplication with \toolname{}}: FaaS containers \emph{madvise} duplicated memory regions to reduce total allocaton.}
    \label{fig:faas_sum}
\end{figure*}

\paragraph{Setup}
To investigate the sharing potential of functions, we profile the memory consumption of four different serverless functions from the SeBS benchmark suite~\cite{sebs}.
We run functions locally on a virtual machine with the help of SeBS's local Docker runtime.
The benchmark suite provides a Docker container with an HTTP server that accepts invocation requests and executes the function’s code upon invocation.
We measure the content similarity between different instances of the same function by running the function with changed inputs on two containers.
The experiments are performed on an x86-64 virtual machine with 24 vCPUs Intel Xeon Platinum 8255C and 48GB of memory.
We used Ubuntu 18.04 with kernel 4.15.18, Docker 20.10.6, Python 3.7.5, and SeBS 1.0.

For profiling, we select workloads representing diverse computation, memory, and I/O requirements:
\begin{itemize}
    \item \textbf{dynamic-html} A lightweight web application that generates dynamic HTML from a template. 
    \item \textbf{thumbnailer} A function that creates thumbnails of uploaded images with the help of the Pillow library.
    \item \textbf{image-recognition} A standard image recognition function, which downloads a pre-trained ResNet-50 model from storage and classifies a image using PyTorch. 
    \item \textbf{dna-visualization} A function that generates a visualization for a given DNA data and stores result in the storage.
\end{itemize}

Figure~\ref{fig:sharing_potential} presents the memory sharing potential of each benchmark.
We include the volatile memory that is not identical, the file-backed memory already shared by OverlayFS,
the identical but not shared anonymous memory, and the identical but not shared file-backed memory.
For all of the benchmarks, the memory that is not shareable accounts for over half of the function memory.
This includes the input, a large part of the memory footprint of small functions.
Most of the file-backed library pages are shared by being page cache enabled in the Docker OverlayFS.
A few libraries are loaded dynamically into memory, resulting in not much shareable file-backed memory.
The ML inference function \textit{image-recognition} shows the greatest memory-sharing potential compared to the others and the greatest overall memory footprint.
There, about 40\% of total memory can be shared, comprised of 27\% and 13\% in the anonymous and file-backed memory, respectively.

We notice that the memory of each benchmark grows soon after a request is received, then drops and remains constant once the request has finished processing. 
This phenomenon raises the question of whether memory deduplication should be applied at the moment when memory consumption reaches its peak.
However, a closer inspection reveals that the request memory causes the increase and the magnitude of the increase is correlated with input data size.
The sharing potential of input data size is minimal as identical inputs are unlikely.

\paragraph{Optimizations}
To understand the potential for sub-page sharing, we select non-identical pages and verify if a sub-page of it would be shareable.
We find that less than 2\% of all pages are 75\% identical, and less than 3\% of all pages are 50\% identical for each function.
Thus, the overhead imposed by deduplicating data units smaller than a memory page %and adding patches
would be much greater than the benefits of sharing.

Furthermore, we inspect the effects of address space layout randomization.
By disabling ASLR, we increase the memory deduplication effect by 5.8 percentage points on average, which validates with another result achieved on a different set of functions~\cite{10.1145/3492321.3524272}.
Therefore, the ASLR is not a significant limitation for the effectiveness of memory deduplication.

\paragraph{Summary}
To summarize, three %(\textit{dynamic-recognition}, \textit{thumbnailer}, and \textit{dna-visualisation}) of the four
of the four benchmark functions have limited deduplication potential because OverlayFS has already shared most of their identical memory through page cache sharing.
On the other hand, the \emph{image-recognition} function has significant memory sharing potential, with roughly
40\% of memory that can be deduplicated.
This result indicates the memory structure of machine learning inference functions makes them good candidates for memory deduplication.
Finally, sub-page level sharing and disabling ASLR have little impact on the memory deduplication of serverless functions.

%-------------------------------------------------------------------------------
\section{User-guided Page Merging}

\label{chapter:design}

We propose User-guided Page Merging (\toolname{}), a new Linux kernel module for memory deduplication, named after the Kernel Same-page Merging (KSM).
\toolname{} supports sharing both anonymous and file-backed memory and is optimized to deduplicate the memory of serverless functions.
\toolname{} retains the classic ideas of hashing and copy-on-write page merging from KSM but replaces content-based page sharing with application hints (Figure~\ref{fig:faas_sum}).
In this section, we outline the design goals and provide a high-level overview of the system, followed by a detailed analysis of \toolname{} components (Sec.~\ref{chapter:implementation}).

\subsection{Design Goals}
\toolname{} fulfills several requirements motivated by the characteristics of serverless workloads.

\textbf{Speed}
Since content-based scanning is not fast enough to discover short-lived pages efficiently, \toolname{} must be able to focus on identical pages in serverless
environments instead of performing random scanning.

\textbf{Focus on user data}
In virtual machines, 63-93\% of shareable pages are part of the page cache~\cite{xen}.
Since page cache sharing is enabled by default in containerized workloads, \toolname{} does not require
specific treatment of I/O devices.

\textbf{Stable Pages}
KSM is looking for stable pages that don't change their contents frequently.
Similarly, \toolname{} looks for memory pages that stay constant across invocations to avoid futile deduplication.
Prior research on serverless functions indicates that at least 76\% of memory pages are the same across invocations with changing inputs, and for a majority of functions, this ratio is larger than 97\%~\cite{10.1145/3445814.3446714}.
Therefore, \toolname{} can assume that consecutive invocations will not change most of the function’s memory pages,
and no additional classification of static pages is needed.

\textbf{Compatibility}
The serverless landscape includes various runtimes and languages.
Therefore, \toolname{} should be compatible with existing systems, and should not require FaaS platforms to modify their internal implementation significantly.
Since \toolname{} is implemented as a kernel module, the deployment requires only an adjustment in the operating system configuration, leaving the FaaS runtime untouched.

\subsection{FaaS with \toolname{}}
The overview of applying \toolname{} to FaaS is presented in Figure~\ref{fig:faas_sum}. 
When a function begins executing in a container and initializes its data structures,
it starts the page sharing process with the system call \texttt{madvise} (\ding{182}).
The provided memory pages are scanned on the first container, and their hashes are inserted into
\toolname{} hash tables.
When \texttt{madvise} is called from a second function container (\ding{183}), the call will detect pages with identical content.
Then, the memory pages of the second container are replaced with references to pages of the first container with an update of OS resources (\ding{184}) and flushing hardware caches (\ding{185}).
The deduplication takes place only on the first cold invocation in a container.
For all consecutive executions, functions immediately benefit from the shared memory and incur no CPU overhead normally associated with memory deduplication.

\begin{figure}[t!]
  \centering
  \includegraphics[width=\linewidth]{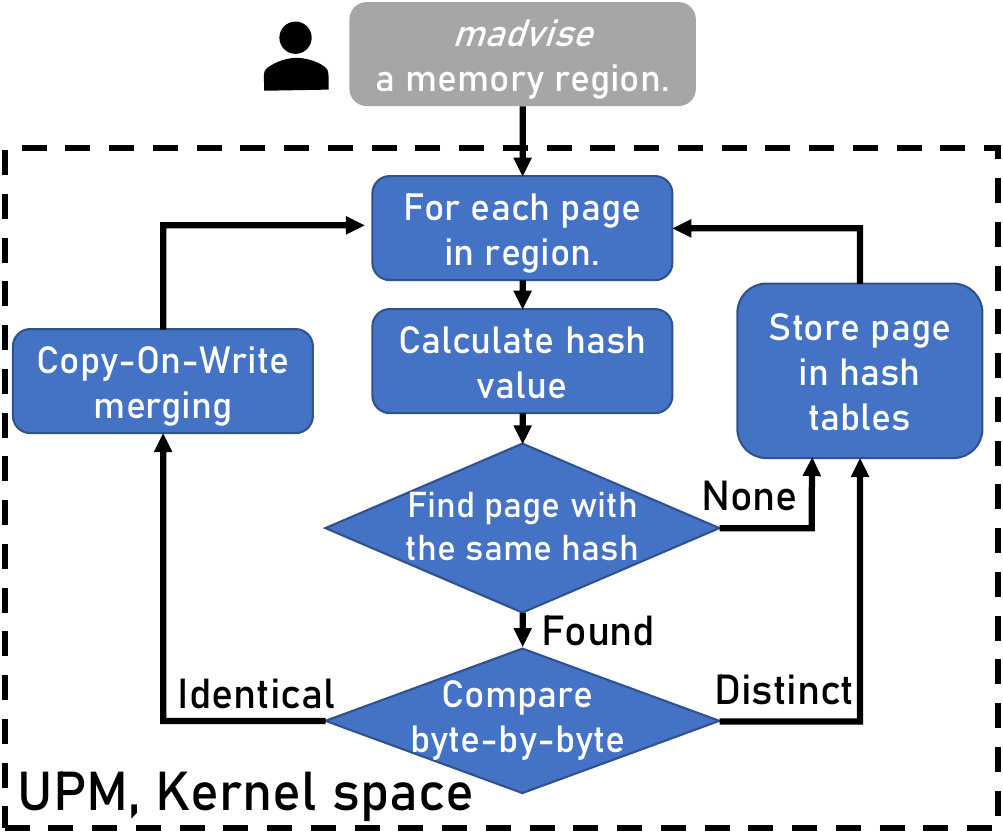}
  \caption{The main deduplication algorithm of \toolname{}.}
  \label{fig:simple_flow_chart}
\end{figure}

\subsection{Guided Page Deduplication}
The distinguishing characteristic of \toolname{} is that we delegate determining sharing candidates to the user.
\toolname{} does not scan the entire system memory.
We believe users have adequate knowledge of application-specific data structures and can determine whether or not a memory region is a good candidate for sharing.
The location of large and constant memory regions can be obtained with the help of profiling and analyzing container snapshots~\cite{10.1145/3445814.3446714},
a process that can be applied automatically to function invocations in the cloud.
We let the application provide \emph{hints}, a well-established technique previously used to optimize file system, networking, memory, and prefetching~\cite{10.1145/224057.224064}.
To that end, we use the existing \emph{madvise} system call:
\begin{verbatim}
int madvise(
  void *addr, size_t length, int advice
);
\end{verbatim}
This system call lets the kernel know how it is expected to handle a given memory area: \texttt{addr} is the virtual address in the memory, \texttt{length} is the size of the area, and \texttt{advice} indicates which kind of advice the program would like to give.
Consequently, \toolname{} can only share memory regions explicitly allocated by the user with known addresses.

The main idea of the algorithm is shown in Figure~\ref{fig:simple_flow_chart}.
When \toolname{} is recommended to deduplicate a specific memory region, \toolname{} calculates hash
values for each page in the region and searches for previously reported pages with the same
hash value.
Then, a byte-by-byte comparison is applied to verify that pages are truly
identical when the hash value matches.
Afterward, the pages are merged in copy-on-write semantics.
If any process attempts to modify the page's contents in the future, the operating system will replace the single copy of the deduplicated page with two instances.

\toolname{} incurs no overhead unless users indicate deduplication candidates.
Thus, \toolname{} does not require kernel-level configuration and switching.
It will remain inactive by default if no calls to the \code{madvise} system routine are made.

\subsection{Concurrency}
Unlike KSM, which scans and shares pages in the background using a single kernel thread, \toolname{} shares pages synchronously, which means the function will be blocked by the \texttt{madvise} system call and will only continue to execute when the system call returns.
Alternatively, the request can be processed asynchronously to avoid adding overhead to the function runtime.
For example, deduplication can be executed in the background of the function invocation, or cloud providers can schedule this task between invocations.
%The second reason is that we should not employ any background threads that use CPU clocks because this goes against the principle that serverless system customers only pay for what they use.

\toolname{} supports safe access to the pages being \emph{madvised} and ensures consistency. Multiple containers can request \toolname{} deduplication simultaneously.
All accesses to data structures are protected with kernel spinlocks.
We use write-protection and byte-to-byte comparison to ensure that the page being considered is not modified while the deduplication takes place.

\begin{figure*}[th!]
  \centering
  \includegraphics[width=\linewidth]{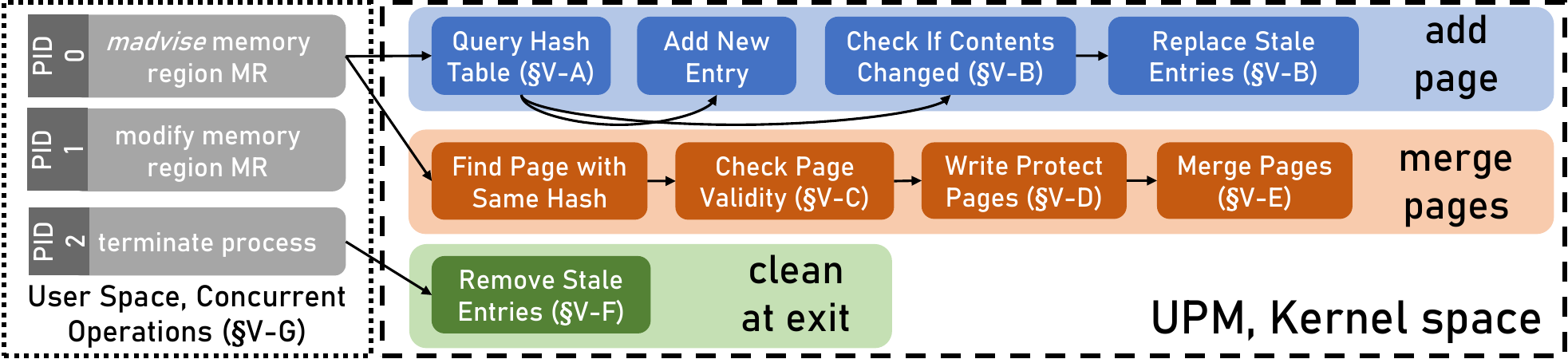}
  \caption{\toolname{} supports concurrent operations on memory pages and exposes methods for managing deduplication.}
  \label{fig:algorithms_full}
\end{figure*}

\subsection{Security}
\toolname{} extends the security model of KSM and shared page caches with \emph{controlled} sharing.
The hint generation is done entirely on the user side, constituting explicit and opt-in permission for sharing.
Page sharing is limited to data explicitly selected by the user, and other tenants cannot detect the contents of memory pages not selected for deduplication.

\iftr
Since we employ page sharing in \toolname{}, it is possible to detect that sharing of a specific page occurs through a side-channel attack~\cite{10.1145/3167132.3167151,10.1145/1972551.1972552}.
\else
Since we employ page sharing in \toolname{}, it is possible to detect that sharing of a specific page occurs through a side-channel attack~\cite{10.1145/1972551.1972552}.
\fi
However, a potential leak does not provide any information on functions using the same memory pages,
except that such data exists in the system.
Moreover, since serverless systems are black-box and do not allow users to control the function placement,
attackers can only achieve co-location with another tenant through trial and error.

%-------------------------------------------------------------------------------
\section{Implementation Details}

\label{chapter:implementation}

%-------------------------------------------------------------------------------

In this section, we detail the challenges and features of the \toolname{} implementation, answering the following questions:

\begin{itemize}[noitemsep]
  \item What hash algorithm and data structures do we use for \emph{madvised} pages? (Sec.~\ref{sec:hashing})
  \item How does \toolname{} handle \emph{madvise} on the same address, but with different content? (Sec.~\ref{sec:rmap_hash_table})
  \item How does \toolname{} handle page modifications before they are deduplicated?  (Sec.~\ref{sec:check validity})
  \item How does \toolname{} prevent page modifications during \toolname{} deduplication? (Sec.~\ref{sec:write-protect})
  \item How does \toolname{} implement the page merging? (Sec.~\ref{sec:page merging})
  \item How does \toolname{} handle cleaning memory pages when processes exit? (Sec.~\ref{sec:upm_exit})
  \item Does \toolname{} support concurrent access? (Sec.~\ref{sec:concurrency})
\end{itemize}

We present an overview of \toolname{} operations in Figure~\ref{fig:algorithms_full}.

\subsection{Hash Tables} \label{sec:hashing}
In \toolname{}, we search, insert, and delete \emph{madvised} pages,
and we do not need to store and retrieve elements in a specific order.
Therefore, a hash table is the most suitable structure to store the sharable pages
since it enables search and modifications in O(1) time on average.
We reuse the well-tested Linux built-in hash table defined in \code{linux/hashtable.h}.
The built-in hash table fulfills all of our requirements, and we can avoid increasing the code complexity in the kernel space.
This hash table is implemented as an array of linked lists using separate chaining to resolve hash collisions.

Hash table performance depends on the hash algorithm and its load factor,
defined by the ratio of elements to buckets in the table.
First, we use the \code{xxHash} function provided in the Linux kernel because of its high performance.
Then, based on profiling results (\ref{chapter:profiling}), we assume that most serverless functions merge at most 50\% of their memory.
Furthermore, previous work has shown that 90\% of serverless functions consume no more than 400 MB~\cite{254430}.
Therefore, the hash table is set to best perform with \emph{madvised} pages of size up to 200MB.
%
%Furthermore, the size is increased by a coefficient of 1.3, as recommended for hash tables~\cite{size1.3}.
Then, the size is increased by a coefficient of 1.3, a common rule of thumb for hash tables.
For page sizes of 4kB, the hash table is set to have $\frac{200MB}{4KB} \cdot 1.3$ different buckets.
Since each hash table bucket is just an 8-byte pointer, the hash table adds 520kB of space overhead.
The hash table size does not pose a challenge even for larger memory allocations, such as in machine learning models;
assuming \emph{madvised} pages of up to 2GB increases the static table size only to 5 MB.

Each bucket entry contains the virtual address (8 bytes), pointers to the
page and memory descriptors (8 bytes each), linked list pointers (16 bytes),
and the actual hash value to resolve conflicts (8 bytes).
Thus, each \emph{madvised} page requires 48 bytes, and the space overhead is $1.17\%$ of deduplicated memory.
%
% Why 3120? 200 MB/4KB entries * 1.3, with 48 bytes creates 3120 kB. 
This overhead per function is at most 3120kB,
%
%This maximum memory size is reached only in the pessimistic case that functions \emph{madvise} many distinct memory pages.
and it is reached only in the pessimistic case when function containers \emph{madvise} many unshareable pages.

\subsection{Reversed Hash Tables}
\label{sec:rmap_hash_table}
The same virtual address can be \emph{madvised} multiple times by the user.
As long as the contents of the memory page have not changed, \toolname{} can ignore subsequent calls.
\iftr
To verify if page contents have changed, we borrow from KSM the concept of \emph{reversed} map~\cite{ksm_kernel},
where the usual mapping from hash value to address is reversed.
\else
To verify if page contents have changed, we borrow from KSM the concept of \emph{reversed} map,
where the usual mapping from hash value to address is reversed.
\fi
This hash table uses a virtual address (8 bytes) as an index to identify a page entry, and we store the hash value there (8 bytes).
Virtual addresses alone cannot uniquely identify a page since different processes may use the same address.
Thus, we store the memory descriptor (8 bytes) to identify the process.
By querying the reversed map, we detect if the same process has previously \emph{madvised} this memory page with different contents and replace the stale entry.
Adding the process identifier (PID, 8 bytes) and linked list pointers (16 bytes), the total size of each entry is 48 bytes, leading to the same space overhead as the previous hash table.

\subsection{Memory Page Validity} \label{sec:check validity}
Functions can modify memory pages after reporting them as sharing candidates. 
When \toolname{} finds a memory page with contents identical to a new one, we verify if the page is still in the memory
by checking the present bit in the page table.
Then, we recalculate the hash value to verify that page contents have not changed.

\subsection{Write-protecting Pages} \label{sec:write-protect}
\toolname{} uses copy-on-write to prevent deduplication of pages that have been modified.
When \toolname{} finds two pages with identical contents, it locks them to prevent the operating system from swapping the pages.
To ensure that the contents of the pages are not changed before they are merged, we apply write protection by resetting the corresponding bit in the page table entry.
Since the pages may be modified just after we compare hash values and before we write-protect them, we perform a byte-by-byte comparison of contents.
Adding write protection to a page does not prevent the page from being modified -- rather, write operations on a write-protected page will generate a page fault, and a new physical page will be allocated.
Thus, in this unlikely scenario, no page deduplication is possible, but \toolname{} will find out
that the page has been modified and handle the change correctly.

\subsection{Page Merging} \label{sec:page merging}
\toolname{} shares memory pages in a copy-on-write fashion.
We merge the newly \emph{madvised} page with the page already present in the hash table.
First, we replace its \emph{page frame number} (PFN) in the page table entry (PTE) of the new page,
making the two virtual pages reference the same physical page.
Before changing the PFN, we acquire the page table locks for both pages and flush the cache and the translation lookaside buffer (TLB).
This prevents functions from accessing the new page contents by using its old page frame.

After merging pages, we renew the reverse mapping information in \toolname{} hash tables.
After the pages are merged and marked as write-protected, any write operation will cause a page fault,
and a new page frame will be allocated for the modified page.
Linux kernel is responsible for processing the page faults, and \toolname{} can use this unmodified.

\subsection{Removing Invalid Entries} \label{sec:upm_exit}
The status of a memory page can change after it has been \emph{madvised} to the \toolname{}.
For example, memory pages can be swapped out, freed, or modified, leaving a stale page entry in \toolname{} hash tables.
However, in the serverless context, the lifetime of such stale entries will be limited by the container’s duration.
Furthermore, the vast fraction of stable pages constrains the frequency of stale entries.
Thus, \toolname{} needs only to clear such entries on the function’s exit.

We implement a cleaning function to clear all entries associated with a process when it exits.
We store a flag for each process to mark if it has added entries to the \toolname{}.
When a process that has used \toolname{} exits, the system iterates over all entries in the reversed hash table to find
pages associated with the PID of the process.
While it would be simpler to iterate over the entire virtual memory area of the process, this solution would not be sufficient.
The addresses of freed memory pages are no longer recorded in the process memory descriptor,
and \toolname{} could keep stale entries if it had inspected only pages belonging to the process at the time of exit.

\subsection{Concurrency and Consistency} \label{sec:concurrency}
\toolname{} allows for safe modifications of memory pages through the verification mechanisms outlined in previous sections and the write protection.
When modifications to the page contents are detected, \toolname{} discards this page as a sharing candidate.
The byte-by-byte comparison will discover modifications if the page is written to before applying the write-protection.
Furthermore, we add a page descriptor comparison right before the page merging to detect page faults.
Therefore, the deduplication has no impact on memory safety, and the \emph{madvise} can be conducted concurrently
with other operations.

The majority of operations performed by \toolname{} are reads needed to calculate hashes and perform comparisons.
The read operations of \toolname{} have no influence on the page, except for the merging when it changes the references to physical pages.
However, flushing caches and TLB will force potential concurrent readers to obtain the physical addresses again from the page table.
Thanks to the page table locks acquired during the merge process before flushing caches (Sec.~\ref{sec:page merging}),
we ensure that readers will not be able to retrieve page table entries while \toolname{} conducts the update.
Thus, when other readers finally access the page table, they will obtain the correct reference to a physical page with identical memory content.

%-------------------------------------------------------------------------------
\section{Evaluation}

\label{chapter:evaluation}
To prove that \toolname{} can be used effectively to reduce the memory footprint in FaaS, we evaluate the deduplication with serverless workloads and answer the following questions:

\begin{enumerate}[noitemsep,nosep]
\item Does \toolname{} decrease memory footprint of functions (Sec.~\ref{sec:memory_containers})?
\item How much memory can be saved (Sec.~\ref{sec:memory_system})?
\item How much time overhead does \toolname{} add (Sec.~\ref{sec:overhead_microbenchmark})?
\item How much slower are cold startups in serverless with \toolname{} (Sec.~\ref{sec:overhead_cold})?
\item Which part of the memory deduplication algorithm takes the most time (Sec.~\ref{sec:overhead_analysis})?
\end{enumerate}

\subsection{Evaluation Setup}
We use two systems to evaluate \toolname{}.
First, we use a slower System A with a Nehalem CPU in the 45 nm process.
Then, we use a higher-performance System B with a Skylake CPU in the 14 nm process.
%
%On both systems, the evaluation is performed using the same virtual machine as described in \cref{chapter:profiling}.
On both systems, the evaluation is performed using a virtual machine running a Linux kernel modified with \toolname{}.
The x86-64 virtual machine runs Ubuntu 18.04 with a modified Linux kernel 4.15.18, built with GCC 8.3.0.
We configure \toolname{} to support sharing of up to 2 GB of memory per function.
We use Docker 20.10.6 and serverless functions from the benchmark SeBS~\cite{sebs} using Python 3.7.5.

\textbf{System A}
The system has four Intel Xeon X7550 CPUs, each one with 8 physical cores at 2.00 GHz frequency, and 1 TB of memory.
It runs Debian 10 with kernel 4.19.

\textbf{System B}
The system has two Intel Xeon 4110 CPUs, each one with 8 physical cores running at 2.10GHz frequency, and 125 GB of memory.
It runs Ubuntu 20.04 with kernel 4.15.18.

\subsection{Benchmarks}
\label{chapter:benchmarks}
We select real-world machine learning inference Python functions to conduct the evaluation from the SeBS benchmark:
\emph{image-recognition} that uses the ResNet50 model, and its counterpart \emph{recognition-alexnet}.
We use the CFFI library to expose the \texttt{madvise} function to Python code.
Since the model is not stored directly in a contiguous memory region, we iterate over its components to \emph{advise} all components. 
We achieve this goal without changing any line of code in PyTorch.

Additionally, we use a microbenchmark to characterize the time overhead of memory deduplication.
The function loads different sizes of random data from the \textbf{same} file into memory
and \emph{madvises} the data.
Since all memory pages are distinct, functions can only benefit from inter-process sharing.

\begin{figure}[t]
	\centering
  \subfloat[Image recognition with ResNet and PyTorch.]{
    \includegraphics[width=1.0\linewidth]{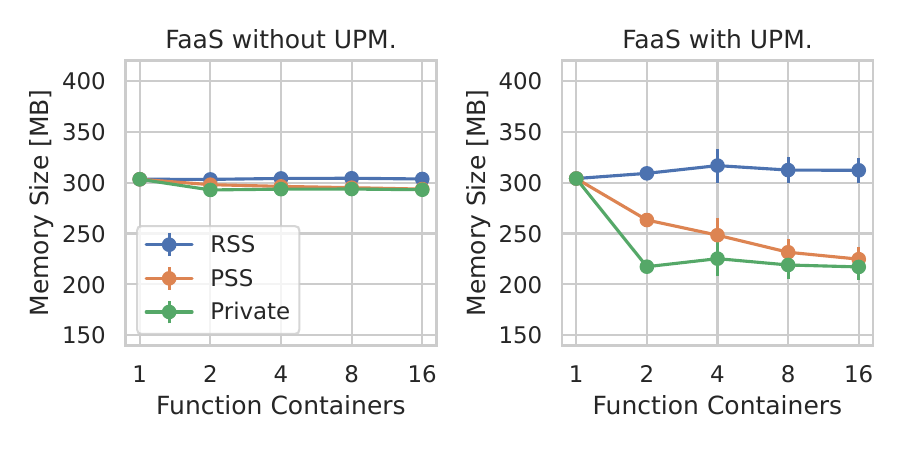}
    \label{fig:concurrent_memory_resnet}
  }
%  \vspace{-1em}
  \hfill
  \subfloat[Image recognition with AlexNet and PyTorch.]{
    \includegraphics[width=1.0\linewidth]{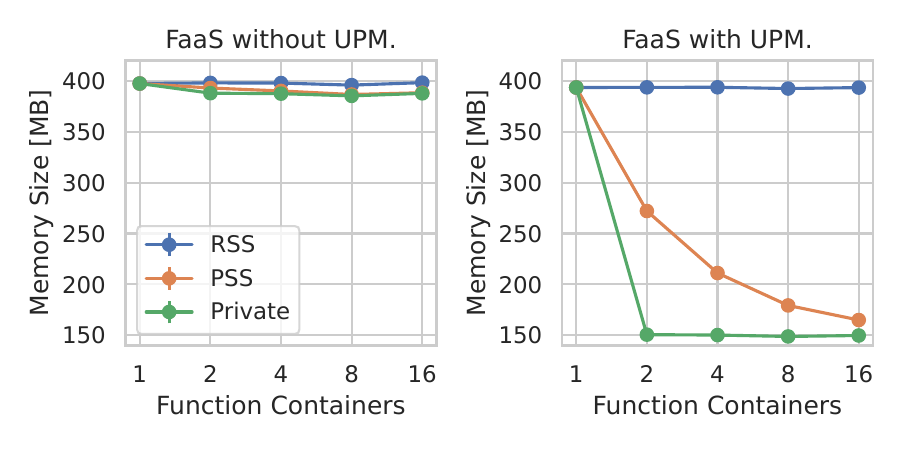}
    \label{fig:concurrent_memory_alexnet}
  }
  %\vspace{-1em}
  \caption{Memory deduplication with \toolname{}: memory consumption of concurrently executing functions.}
  \vspace{-1em}
\label{fig:concurrent_memory}
\end{figure}

\subsection{Memory Usage of Function Containers}
\label{sec:memory_containers}

To evaluate the effectiveness of our memory deduplication approach, we simulate concurrent executions of serverless functions on a local machine
running with the modified Linux kernel (System A).
We deploy each function instance in Docker containers that execute concurrently and can benefit from page deduplication.
For each container, we run five invocations to stabilize memory consumption and measure the memory consumption of the warm function instance.
We measure the resident set size (RSS) defined as the total memory allocation of a process, including zero and shared pages, the private memory size of a process, and the proportional set size (PSS) that adjusts the memory consumption of a process by considering the sharing of memory by $n$ processes:
\setlength{\abovedisplayskip}{2pt}
\setlength{\belowdisplayskip}{2pt}
\begin{equation*}
    PSS = \dfrac{shared \: memory}{n} + private \: memory
\end{equation*}
For $n = 1$, PSS is equal to RSS since no memory is shared.
When increasing the number of processes $n$, the influence of the shared part on the memory consumption decreases.

Figure~\ref{fig:concurrent_memory} presents the measured memory consumption of serverless benchmark functions
for a varying number of concurrently residing warm function instances.
Each data point shows the memory size per function container.

\toolname{} reduces the PSS of each container by 14.1\% with two concurrent containers and 26.4\% with 16 concurrent containers for the ResNet example, and by 29.4\% and 55\% respectively for the AlexNet example.
For AlexNet, \toolname{} reduces the private memory size to around 150 MB across the runs, saving about 250 MB of memory on each container.

\begin{figure}[t]
  \centering
  \includegraphics[scale=0.55]{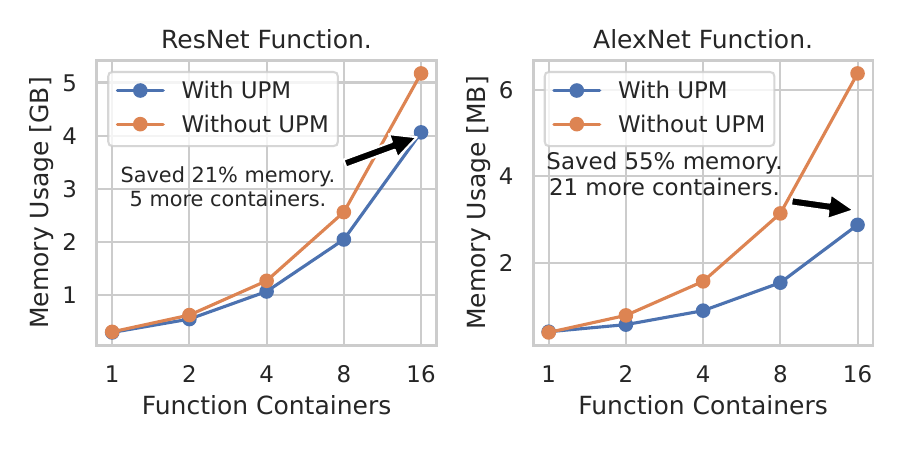}
  \caption{The effects of \toolname{} on memory consumption.}
  \label{fig:memory_increase}
\end{figure}

\subsection{Memory Usage of the System}
\label{sec:memory_system}

%DATA: 
% here we show memory usage of the entire FaaS system, measured as whole OS memory consumption for N containers (but we subtract the OS memory consumption BEFORE the test, so it should show the memory consumption of functions + \toolname{} data structures)

%ResNet - on 16 containers we save 1134 MB and that's 21.4%% of the memory of the FaaS system. Since PSS of ResNet at 16 containers is around 225 MB, we can fit in almost 5 containers
%AlexNet - on 16 containers we save 3585 MB and that's 54.9%% of the memory of the FaaS system. SInce PSS of AlexNet at 16 containers is around 165 MB, we can fit it over 21 more containers

We measure the memory usage of the entire FaaS system, using the same experimental setting as the memory evaluation on containers in \ref{sec:memory_containers}.
We record the memory increase on the system caused by the containers, by executing the Linux command \mbox{\texttt{free -m}} before and after launching the containers and invoking the function five times.
The system's memory usage can better illustrate the memory reduction effect, since it also counts the data structures that \toolname{} creates in the kernel, including the hash tables and the page entries.

Figure \ref{fig:memory_increase} shows that for ResNet, \toolname{} reduces the total memory usage of the system by 20\% with 16 containers running concurrently.
The memory reduction reaches 1134MB.
Given the PSS of around 225MB for a container, five more containers running the same function can be added to the system.
For Alexnet, the reduction is even greater: 55\% with 16 containers, amounting to 3585 MB.
Given the PSS of about 165 MB per container, this means 21 more containers could be added to the system running the same function.

\subsection{Time Overhead of \toolname{}}
\label{sec:overhead_microbenchmark}
\begin{figure}[t]
  \centering
  \includegraphics[scale=0.55]{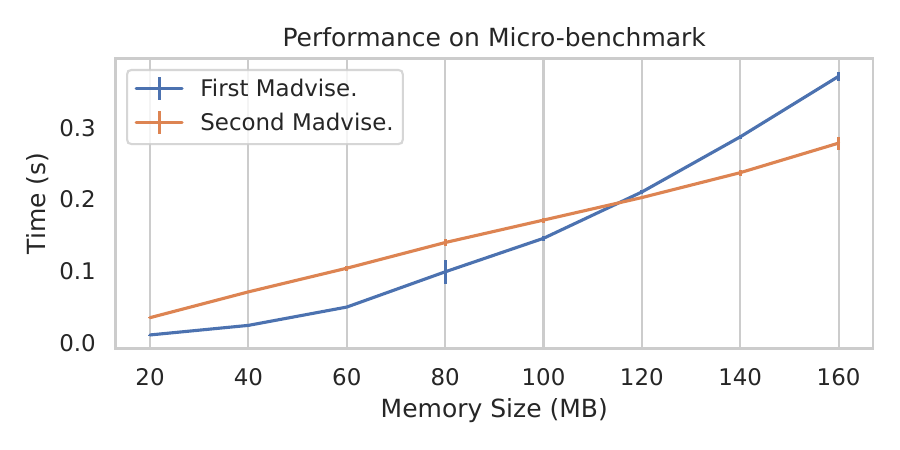}
  \caption{Adding and sharing memory through \emph{madvise}.}
  \label{fig:dummy_memory_sharing}
\end{figure}

We run the microbenchmark on System B to measure the time impact of sharing memory with \toolname{}.
Figure \ref{fig:dummy_memory_sharing} shows the results on the micro-benchmark described in Sec.~\ref{chapter:benchmarks}.
Each data point is the average of 10 runs with the standard deviation.
The first \emph{madvise} call shows the cost of the first launched program that only adds the pages in the hash table without merging any pages.
On the other hand, the second \emph{madvise} call represents the second launched program that also merges the pages.

\begin{figure}[t!]
	\centering
  \subfloat[Image recognition with ResNet and PyTorch.]{
    \includegraphics[width=1.0\linewidth]{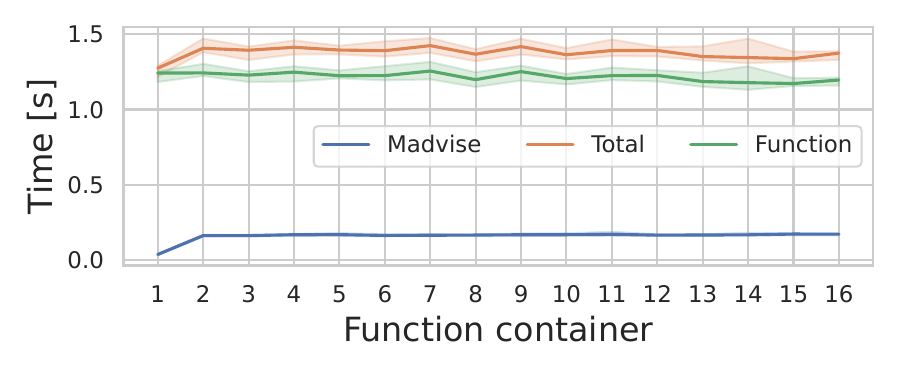}
    \label{fig:time_resnet}
  }
%  \vspace{-1em}
  \hfill
  \subfloat[Image recognition with Alexnet and PyTorch.]{
    \includegraphics[width=1.0\linewidth]{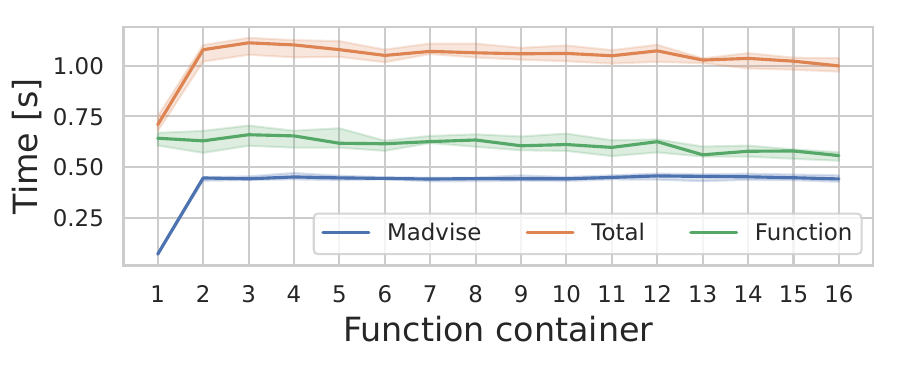}
    \label{fig:time_alexnet}
  }
  %\vspace{-1em}
  \caption{Time overhead of memory deduplication on first cold invocation, reported medians with parametric 95\% CIs.}
  \vspace{-1em}
\label{fig:time_cost}
\end{figure}

\subsection{Time Overhead on Cold Startups}
\label{sec:overhead_cold}
To test using \toolname{} in a FaaS system, we assume users call the \emph{advice} on function data once, in the first invocation of the function during the cold initialization.
In the following experiments, which we repeat ten times, each function is invoked once to measure the time of the \texttt{madvise} system call, and the functions are invoked in a cold container.
We assume the most pessimistic scenario for the performance of \toolname{}, where deduplication is conducted synchronously and the user function blocks until it is finished.

We launch 16 containers running the function and invoke them one by one on System B.
The average and the standard deviation of the time cost of the invocation of each container are shown in Figure \ref{fig:time_cost}.
The figure shows three lines: total time, the time spent on the \texttt{madvise} system call, and the time spent for the function call excluding the \code{madvise} system call.

The overhead introduced by deduplication is equal on average to 12\% and 42\% of function time on ResNet and Alexnet, respectively.
The leap after the first container is explained from page sharing beginning from the second container.
However, this overhead applies to the first cold invocation, only once per the entire container lifetime.
All future warm invocations do not call the \code{madvice} function again.

\begin{table}\centering
%\ra{1.3}
%\setlength{\tabcolsep}{1.5pt}
\begin{adjustbox}{max width=\linewidth}
\begin{tabular}{lcc@{}}\toprule
    \textbf{Functionality}           & \textbf{Sharing}   &   \textbf{Sharing \& Merging} \\
    \hline
    Search in Hash Table    & 4.4\%        &   61.4\%  \\
    Calculate Hash          & 32.5\%        &   19.6\%  \\ 
    Search in Reversed Hash Table          & 51.2\%         &   10.1\%  \\ 
    Merge Pages             & 0\%         &   6.3\%   \\
    Add Page to Hash Table  & 9.4\%         &   0.9\%   \\
    Spin Locks              & 5.5\%         &   2.1\%   \\
\bottomrule
\end{tabular}
\end{adjustbox}
\caption{Distribution of \toolname{} time across the most important parts of the system, for first function (\textbf{Sharing}) and consecutive functions (\textbf{Sharing \& Merging}).
  Components are responsible for 96.2\% and 91.7\% of total runtime of \emph{madvise}. Time of spin locks is included in other components.
}
\label{tab:applications_evaluation}
\end{table}

\subsection{Time Overhead Distribution}
\label{sec:overhead_analysis}
To analyze the sources of the time overhead \toolname{} introduces, we profile the kernel code
and measure the execution time of the \emph{image-recognition} function on System A.
In this function, we \emph{madvise} approximately 100MB of model memory.
Table~\ref{tab:applications_evaluation} presents the breakdown of timing data across most time-consuming functionalities, including the time needed to acquire the spin locks.
The main source of CPU overhead in \texttt{madvise} is iterating over hash table entries.
This is not surprising as \toolname{} has to compare each memory page in the shared region with existing hash table entries. 
One-fifth of the time is spent on hashing memory pages and is limited primarily by the DRAM bandwidth.
We observe that the overheads of using spinlocks when modifying hash tables are relatively low during normal execution.
%

%-------------------------------------------------------------------------------
\section{Discussion}

\label{chapter:conclusions}
In this work, we introduce \toolname{}, a memory deduplication system designed for serverless workloads.
In the following, we discuss the integration of \toolname{} with cloud systems
and analyze how memory sharing can be improved through the synergy with other cloud and FaaS technologies.

\textbf{When to deduplicate?}
We evaluate the pessimistic scenario of deduplication occurring on the critical path to highlight the overheads of our system (Sec.~\ref{sec:overhead_cold}).
In practical deployments, the deduplication will be moved from the critical path, either to a background thread executing or to the cloud operator side.
In the former, users pay the CPU cost of reducing memory consumption in parallel with the function workload.
In the second scenario, users only annotate memory regions at the first invocation,
and the cloud operator triggers deduplication logic in the FaaS runtime.
This scenario comes with no overhead or costs to the user.

\textbf{Who benefits from deduplication?}
In the classic setup of memory deduplication, page sharing is entirely transparent, and users are unaware when their virtual machines share data in the background.
All memory savings benefit the cloud provider directly, and users keep paying the same price for requested resources.
On the other hand, in the serverless setup, users must take the initiative to enable page sharing,
and they can expect performance overheads of deduplication.
Therefore, to incentivize users, cloud providers should allow users to participate in the benefits of saving memory resources.
Since the billing models of FaaS always include a memory component, cloud providers can adjust the price by subtracting memory savings from the requested memory allocation.

\textbf{How can the sharing effectiveness of \toolname{} be further improved?}
The application of \toolname{} opens new directions for automatically detecting memory likely to be shareable.
In many languages, initialization and heap construction can be done at build time~\cite{10.1145/3360610},
allowing static determination of sharing candidates.
Furthermore, serverless functions consist of small codebases that are executed thousands of times
in the cloud.
Such processes can be easily profiled to automatically locate memory pages with identical and constant contents.
%that stay constant across invocations.
%
Similarly to virtual machine fingerprinting and co-location techniques~\cite{10.1145/1618525.1618529,10.1145/1989493.1989554}, containers with sharing potential can be migrated and co-located on a single machine.
%
%These containers could be migrated and co-located on a single machine
%with other instances determined to have a substantial inter-function sharing potential.

\section{Conclusions}
We propose and implement User-guided Page Merging (\toolname{}), a novel memory deduplication
approach optimized for FaaS systems.
\toolname{} improves scanning-based deduplication with user-provided hints to enable deduplication on serverless running for seconds and not hours. 
Furthermore, \toolname{} does not require modifying serverless runtimes,
works for all processes handling serverless workloads, regardless of their language and runtime,
and extends security guarantees of KSM with controlled sharing.
In an evaluation with machine learning inference, we show \toolname{} can reduce system memory utilization
by more than half, and allows FaaS servers to handle more containers without increasing hardware resources.

%\newpage

\section*{Acknowledgment}

  \begin{wrapfigure}{r}{.13\linewidth} %\columnwidth}
    \includegraphics[width=.13\columnwidth]{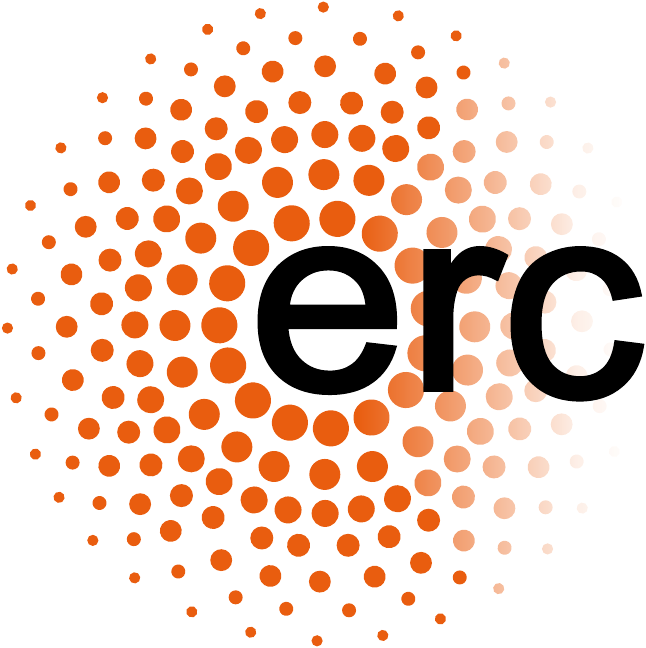}
  \end{wrapfigure}

This project has received funding from the European
Research Council (ERC) under the European Union’s
Horizon 2020 program (grant agreement PSAP, No. 101002047),
and EuroHPC-JU funding under grant agreements DEEP-SEA, No.
95560 and RED-SEA, No. 955776).
We would like to thank Jike Song for providing valuable advice
and support.
We would also like to
thank the Swiss National Supercomputing Centre (CSCS) for
providing us with access to their HPC system Ault. 

\iftr

  \bibliographystyle{IEEEtran}
  \bibliography{paper}

\else

  \setlength{\parskip}{0pt}

  {
    %\ssmall
    %\scriptsize
    \footnotesize
    \setlength{\parskip}{0pt}
    \setlength{\bibsep}{0pt plus 1ex}
    \bibliographystyle{IEEEtran}
    %\textsf{
    
    %\bibliography{paper.short}

    \bibliography{paper}
    %}
    %\textsf{\bibliography{paper}}
  }

\fi

%%%%%%%%%%%%%%%%%%%%%%%%%%%%%%%%%%%%%%%%%%%%%%%%%%%%%%%%%%%%%%%%%%%%%%%%%%%%%%%%
\end{document}